\documentclass[sigconf]{acmart}
\usepackage[utf8]{inputenc}
\usepackage{graphicx}
\usepackage{caption}
\usepackage{subcaption}
\usepackage{amsmath}
\usepackage{hyperref}
\hypersetup{colorlinks=true,linkcolor=blue,filecolor=magenta,
    urlcolor=cyan,pdftitle={Overleaf Example},pdfpagemode=FullScreen}
\usepackage{soul}
 
\AtBeginDocument{%
  \providecommand\BibTeX{{%
    \normalfont B\kern-0.5em{\scshape i\kern-0.25em b}\kern-0.8em\TeX}}}

\setcopyright{acmcopyright}
\copyrightyear{2023}
\acmYear{2023}
\acmDOI{XXXXXXX.XXXXXXX}

\acmConference[CACM]{Make sure to enter the correct
  conference title from your rights confirmation email}{March,
  2023}{USA}
\acmPrice{15.00}
\acmISBN{978-1-4503-XXXX-X/18/06}

\title{Visualizing Progress in Broadening Participation in Computing: The Value of Context}

\author{Valerie Barr}
\affiliation{
\institution{Bard College}
\city{Annandale-on-Hudson}
\state{NY}
  \country{USA}
}\email{vbarr@bard.edu}

\author{Carla E.\ Brodley}
\affiliation{
  \institution{Northeastern University}
  \city{Boston}
  \state{MA}
  \country{USA}
}
\email{c.brodley@northeastern.edu}

\author{Manuel P\'erez-Qui\~nones }
\affiliation{
\institution{University of North Carolina at Charlotte}
\city{Charlotte}
\state{NC}
  \country{USA}
}\email{perez.quinones@uncc.edu}

\date{\today}

\begin{document}

\begin{abstract}
Concerns about representation in computing within the U.S. have driven numerous activities to broaden participation.  Assessment of the impact of these efforts and, indeed, a clear assessment of the actual “problem” being addressed are limited by the nature of the most common data analysis which looks at the representation of each population as a percentage of the number of students graduating with a degree in computing.  This use of a single metric cannot adequately assess the impact of broadening participation efforts.  First, this approach fails to account for changing demographics of the undergraduate population in terms of overall numbers and relative proportion of the Federally designated gender, race, and ethnicity groupings. A second issue is that the  majority of literature on broadening participation in computing (BPC) reports data on gender {\bf or} on race/ethnicity, omitting data on  students' intersectional identities. This leads to an incorrect understanding of both the data and the challenges we face as a field.  In this paper we present several different approaches to tracking the impact of BPC efforts.  We make three recommendations: 1) cohort-based analysis should be used to accurately show student engagement in computing; 2) the field as a whole needs to adopt the norm of always reporting intersectional data; 3) university demographic context matters when looking at how well a CS department is doing to broaden participation in computing, including longitudinal analysis of university demographic shifts that impact the local demographics of computing. 
  
\end{abstract}
\maketitle


\section{Introduction}

Concerns about representation in computing within the U.S. have driven numerous activities to broaden participation.  Assessment of the impact of these efforts and, indeed, a clear assessment of the actual “problem” being addressed are limited by the nature of the most common data analysis which looks at the representation of each population as a percentage of the number of students graduating with a degree in computing. In this paper we call this method the {\em standard analysis} (see Figure \ref{standard-rep} for an example of the standard analysis). As pointed out by Barr \cite{Barr-Inroads-2018}, the standard analysis of Computer Science (CS) degree data does not take into account the changing demographics of the undergraduate population in terms of overall numbers and relative proportion of the Federally designated gender, race, and ethnicity groupings.\footnote{We examine visualizations using race/ethnicity and gender in this paper but note that there are other categories of diversity.} While it does give an indication of a student's experience walking into a classroom, and is somewhat reflective of overall current demographics and historic marginalization, a new framework is necessary to evaluate longitudinal change for each demographic group.  A second issue we observe is that the  majority of literature on broadening participation in computing (BPC) reports data on gender {\bf or} on race/ethnicity, omitting data on  students' intersectional identities. This leads to an incorrect understanding of both the data and the challenges we face as a field by using a single-axis of analysis~{\cite{Crenshaw1989}} at a time (gender or race/ethnicity). When used as a framework, intersectional analysis, a term coined by Crenshaw~{\cite{Crenshaw1989}}, allows us to expose the multidimentionality of experiences that Black Women, in Crenshaw's work for example, experience in everyday life. In computing, a number of researchers \mbox{\cite{Rankin2019,Rankin2021,Thomas2018,Ovalle2023,Warner2021,Trauth2012}} have been exploring intersectionality as a framework of analysis for exploring broadening participation in computing.

We argue that, in order to truly assess the effectiveness of curricular, pedagogic, and institutional interventions, we should use multiple data analysis methods, each of which presents a different perspective on the situation and the improvements achieved.  These different analyses allow us to distinguish between the experience a student may have walking into a CS classroom at a particular institution relative to their experience walking into a non-CS classroom, the extent to which the CS department at institution X is representative of the demographics of students across all disciplines at X, and the extent to which CS as a field is attracting and retaining students of different identities.

In this paper, we discuss the challenges of using the {\em standard analysis} to understand representation and, in particular, to understand longitudinal data. We present a series of visualizations that analyze intersectional representation in computing in the context of university demographics across all degrees. We then turn to examine how well the information-based metrics of diversity used in many other disciplines can serve to analyze demographic diversity in computing.  We conclude with a strong recommendation that the BPC community rethink how to represent demographic graduation data in computing and we discuss the limitations of using graduation data for assessing the impact of BPC activities.  Indeed, graduation data ignores issues of retention, persistence, belonging, and the institutional changes needed to attract students to computing independent of prior computing experience in high school.

\section{Cohort Analysis of Longitudinal Degree Data}
\label{sec:cohort}

\begin{figure}
    \centering
    \includegraphics[width=3.0in]{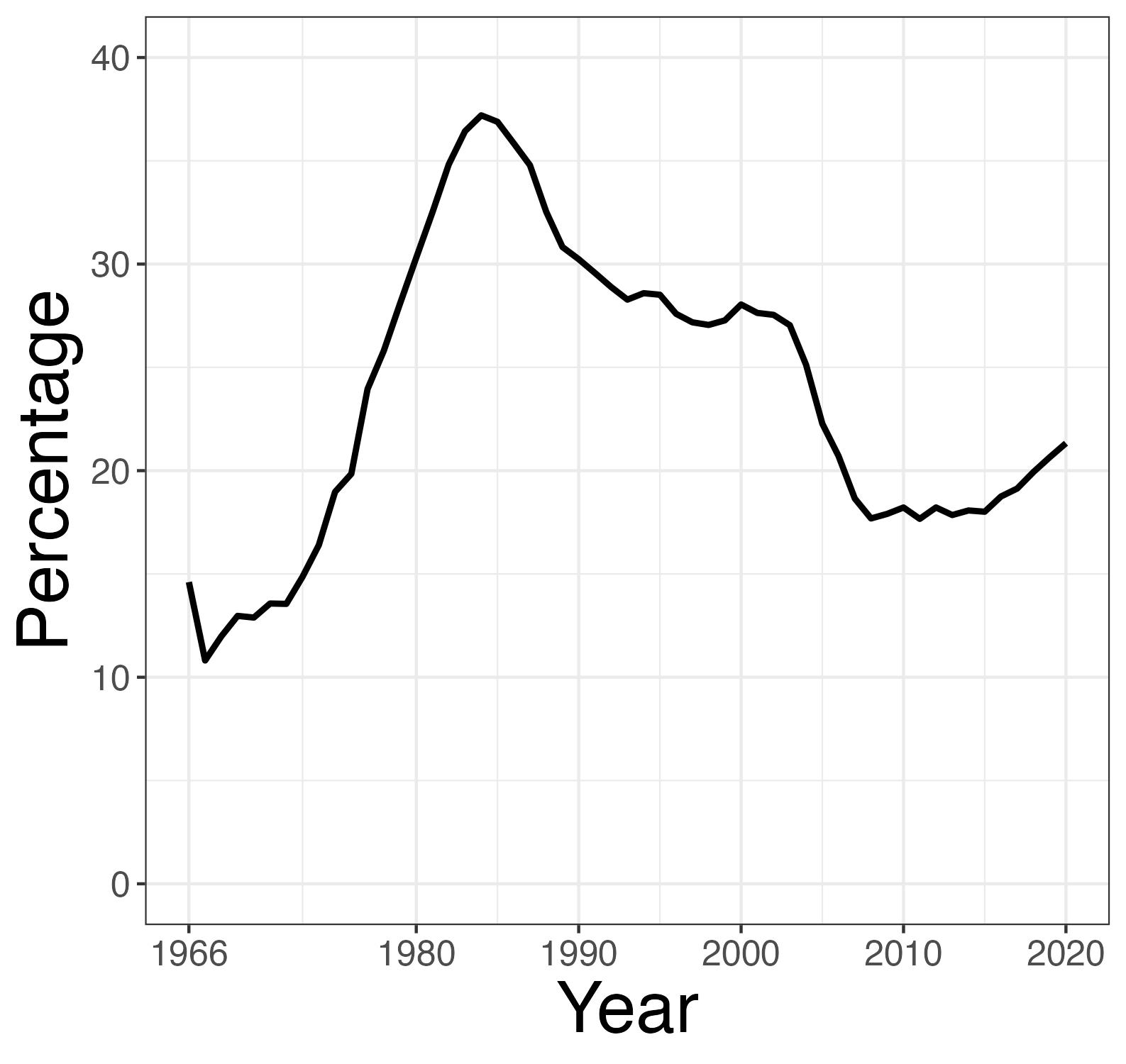}
    \caption{Percentage of total CS degrees earned by women, as an example of the standard analysis}.
    \label{standard-rep}
\end{figure}

Discussion of diversity in computing typically looks at the degrees earned by subgroups as a percentage of the whole.  For example, women's participation in computing is typically based on examination of the percentage of total CS degrees that are earned by women each year, as shown in Figure~\ref{standard-rep}.\footnote{Note that all data presented in this paper are based on U.S. institutions.} The data in this graph are from the Integrated Postsecondary Education Data System (IPEDS) Completions data set.\footnote{https://ncsesdata.nsf.gov/home} IPEDS data is divided by Classification of Instructional Programs (CIP) codes. Computer Science data can be found in Federal CIP code 11. However, for some universities CIP-11 includes Information Technology (IT) and other similarly named programs, so care must be taken in analyzing results because IT degrees are often more diverse than are CS degrees.

\begin{figure}
    \centering
    \includegraphics[width=3in]{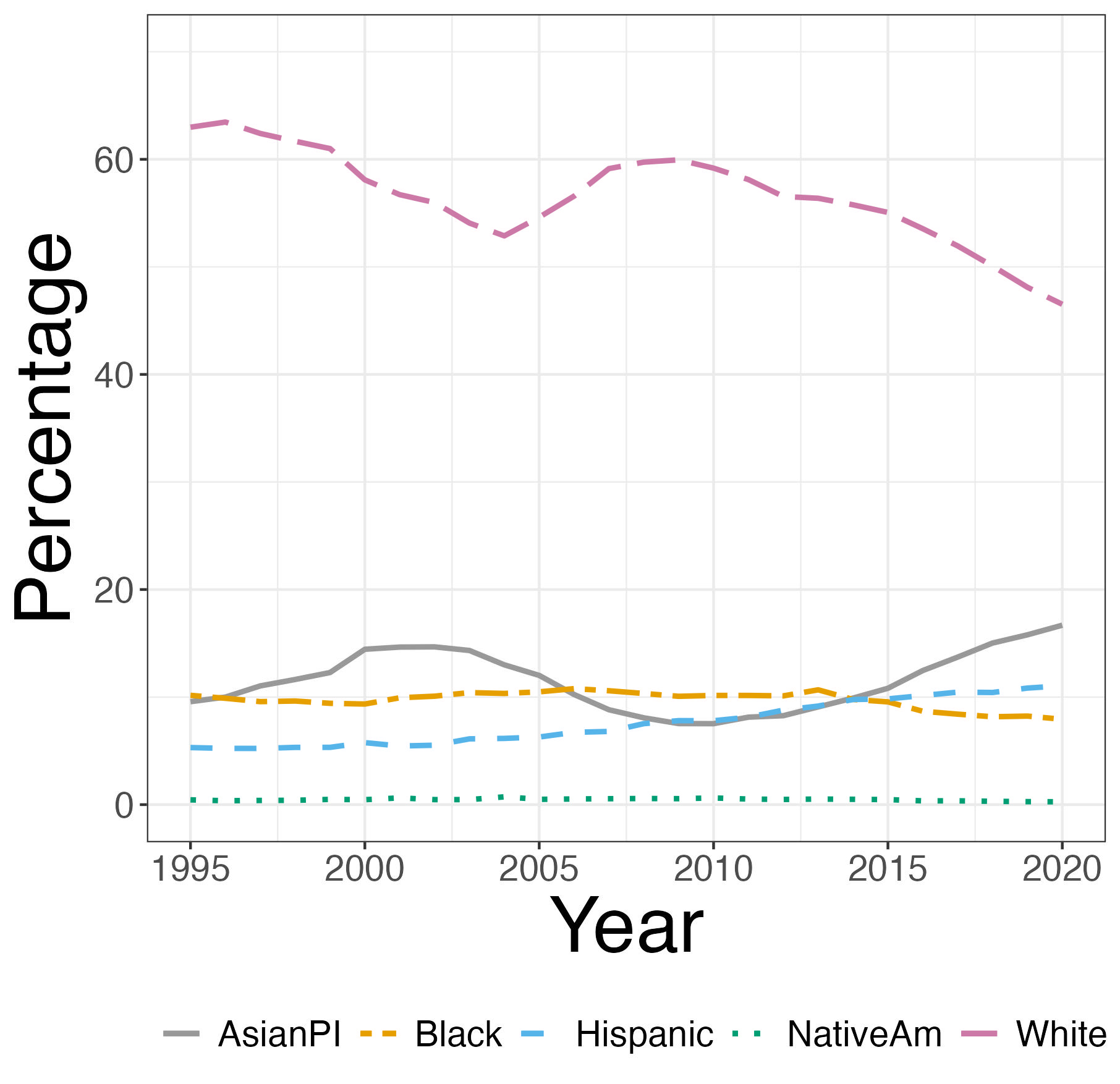}
    \caption{Distribution of CS degrees across IPEDS race and ethnicity categories.}
    \label{standard-race-ethnicity}
\end{figure}

As we discuss in detail below, the standard analysis (see Figure~{\ref{standard-rep}}) does not support an accurate analysis of longitudinal trends.  It does, however, provide a realistic picture of the experience an individual student has as they go through their CS studies.  For example, in 2020, 21\% of CS degrees were awarded to women, which means that a woman CS major walking into a CS classroom of 100 CS seniors would on average see 20 other women (in addition to themself).  Similarly, the standard analysis of CS degrees as reported in IPEDS by race and ethnicity categories, shown in Figure~\ref{standard-race-ethnicity}, illustrates that, in 2020, a Black CS major in a class of 100 students would see 8 other Black students, 57 white students etc.  

The standard analysis is otherwise problematic, particularly for longitudinal analysis of change over time.  It does not, for example, account for significant demographic changes that have taken place in the college-going population over time.   In 1966 (leftmost data point in Figure~\ref{standard-rep}),  women made up 42\% of the U.S.\ undergraduate population, but by 2020, (rightmost data point in the figure) women made up 57\% of the U.S.\ undergraduate population. Thus, although Figure~\ref{standard-rep} gives us men and women's participation relative to each other, it does not show shifts in interest by either group over time.  That is to say, what looks like a sudden drop in women's degrees in the 1980's might actually be a sudden increase in interest by men while women's interest stayed steady.  The relative nature of the data presentation obscures the actual interest in CS indicated by the data.

\begin{figure}
    \centering
    \includegraphics[width=3in]{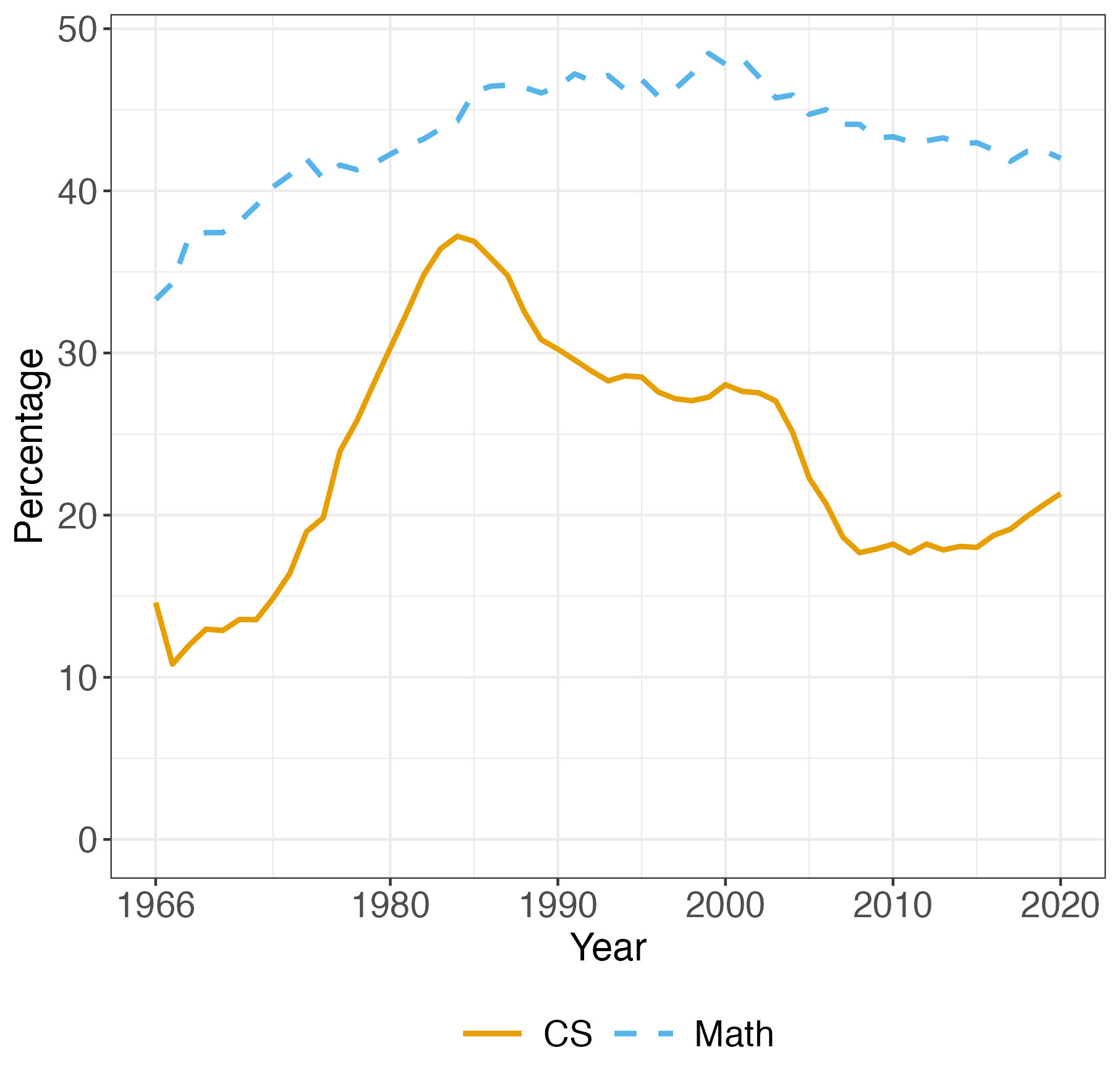}
    \caption{Women's CS and Math degrees as percentage of discipline degrees.}
    \label{WomenCSMathOutOfTotals}
\end{figure}

\begin{figure}
    \centering
    \includegraphics[width=3in]{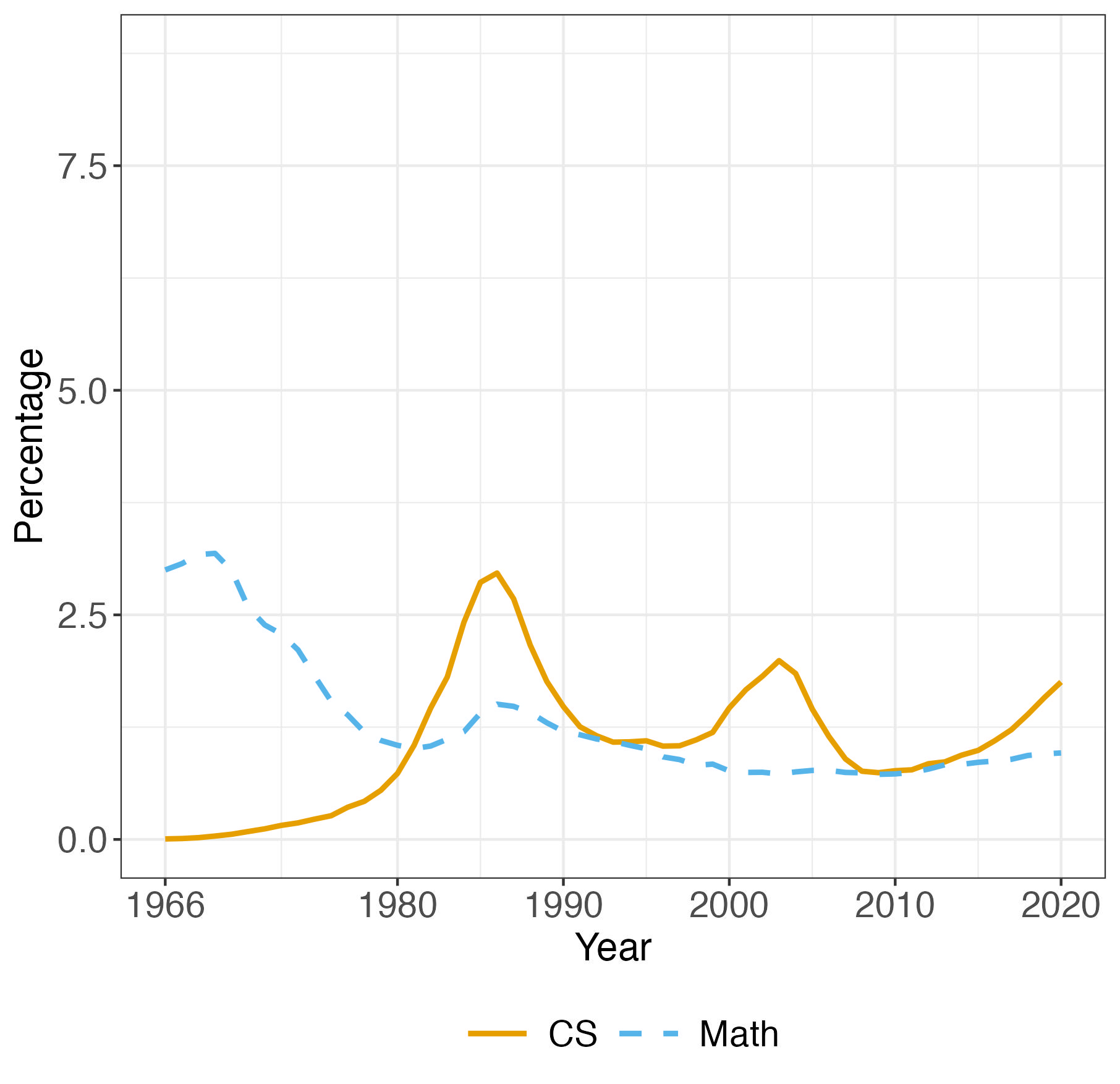}
    \caption{Women's CS and Math degrees as percentage of women's degrees.}
    \label{WomenCSMathOutOfWomen}
\end{figure}

\begin{figure}
    \centering
    \includegraphics[width=3in]{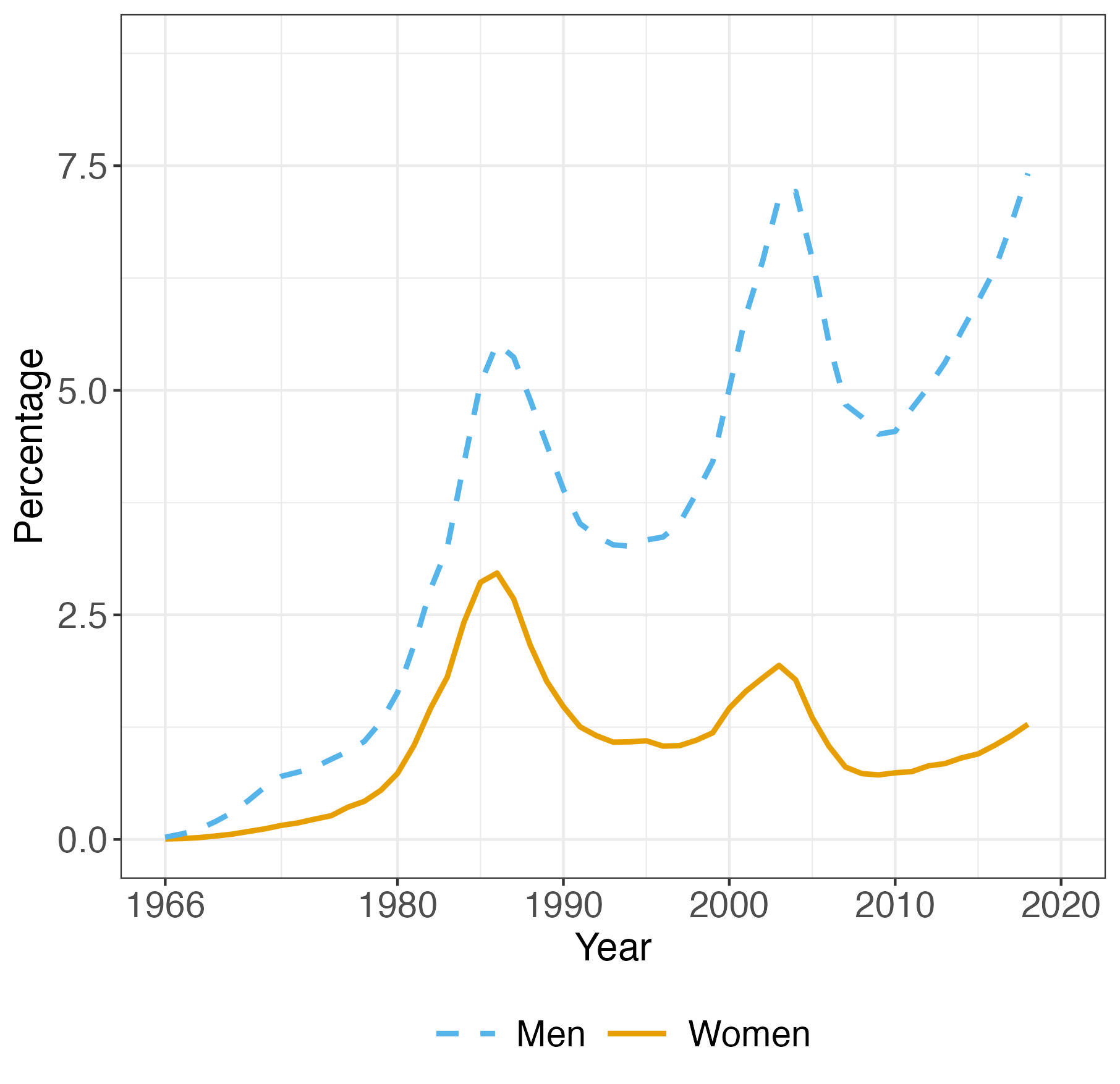}
    \caption{Women's CS and Men's CS as percentage of each cohort's total degrees.}
    \label{WomenMenPercentCSCompare}
\end{figure}

We see this phenomenon clearly when we use the standard analysis to compare women's CS degrees to women's Math degrees.  Figure~\ref{WomenCSMathOutOfTotals} shows the percentage of CS degrees and the percentage of Math degrees that were earned by women during 1966-2020.  From this graph we might conclude that women study math at a significantly higher rate than they study CS because they earn a much higher percentage of math degrees than they do of CS degrees.  Yet this view of the data completely hides the extent to which students do or do not study each field, making a relative comparison inappropriate and erroneous.  Figure~\ref{WomenCSMathOutOfWomen} changes the computation, showing women's CS degrees and women's Math degrees, each as a percentage of women's degrees across {\em all} fields. Figure~\ref{WomenCSMathOutOfWomen} is an accurate representation of the extent to which each field attracts women, independent of how many men study the field. The story told by Figure~\ref{WomenCSMathOutOfWomen} is quite different than that told by Figure~\ref{WomenCSMathOutOfTotals}:  Figure~\ref{WomenCSMathOutOfWomen} shows that women's pursuit of Math degrees fell off by 1980 and is currently below women's pursuit of CS degrees.  Figure~\ref{WomenCSMathOutOfTotals} cannot correctly show this reality because it is distorted by the fact that men study Math at a much lower rate than they study CS, making women's interest in Math {\it appear} higher than it actually is. Figure~\ref{WomenMenPercentCSCompare} shows men and women's CS degrees as a percentage of all men and women graduates and makes clear that, despite increased interest in the field by both groups, men's pursuit of CS degrees has increased far more rapidly than has women's.

\begin{figure}
    \centering
    \includegraphics[width=3in]{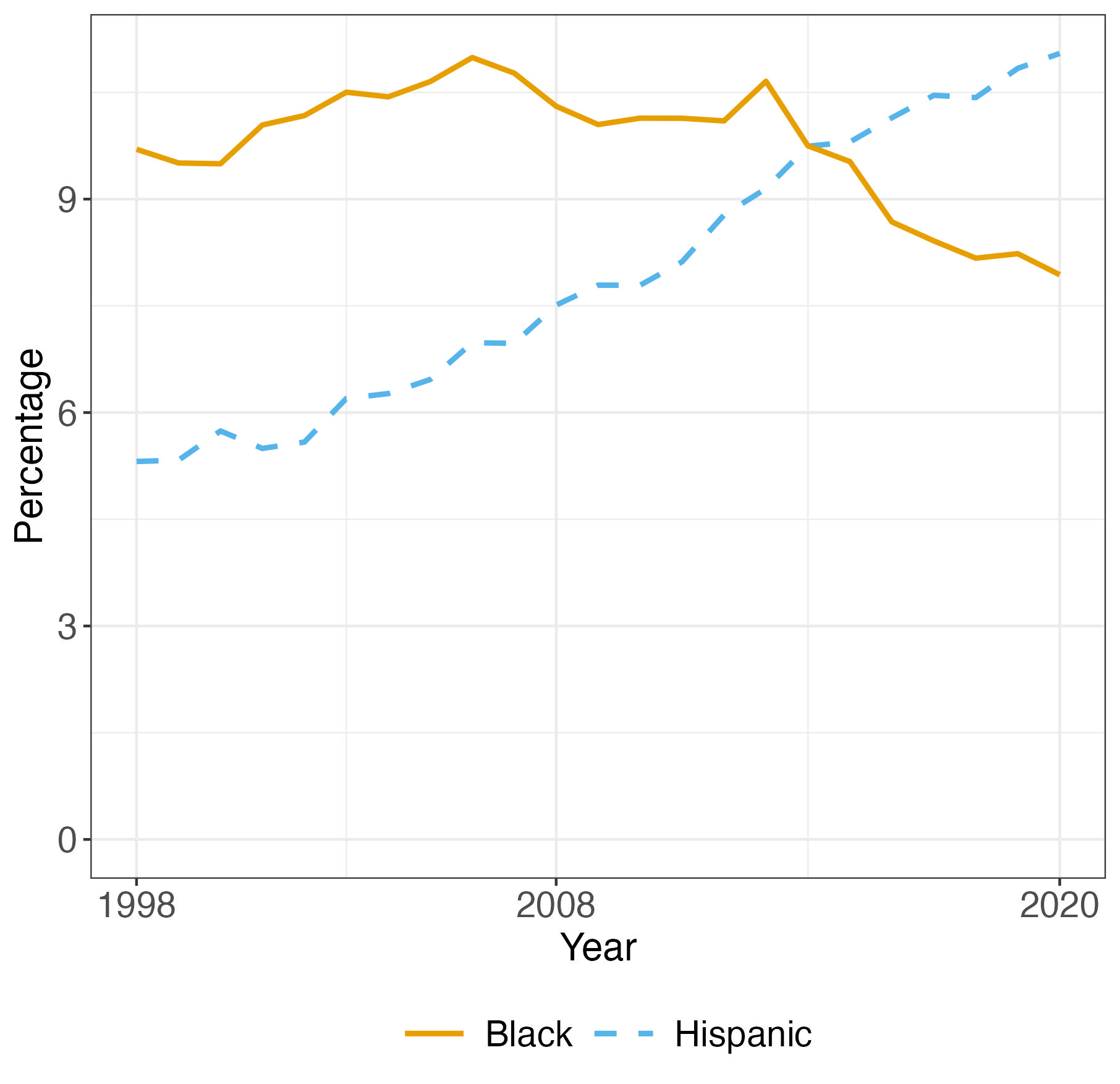}
    \caption{Hispanic and Black CS degrees, percentage of all CS degrees (standard analysis).}
    \label{HispanicBlackCSOutOfCS}
\end{figure}

\begin{figure}
    \centering
    \includegraphics[width=3in]{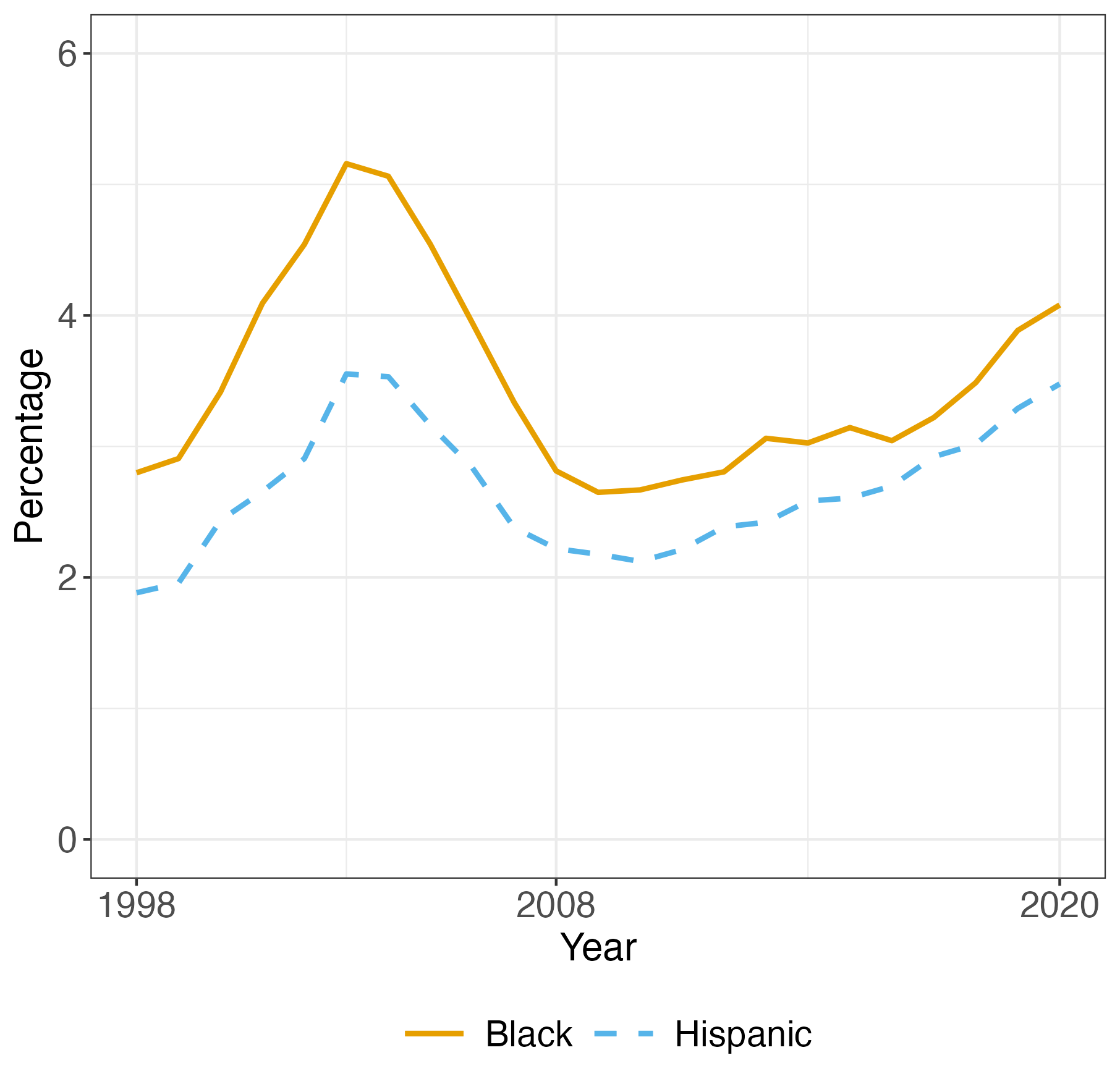}
    \caption{Hispanic and Black CS degrees, percentage of population cohort.}    \label{HispanicBlackCSOutOfCohort}
\end{figure}

As another example of the importance of cohort analysis, we examine CS degrees earned by Hispanic and Black students.  Figure~\ref{HispanicBlackCSOutOfCS} shows the standard analysis with Hispanic CS degrees and Black CS degrees as a percentage of total CS degrees.  One might conclude from this figure that there was a sharp increase in participation in CS on the part of Hispanic students with a concomitant decrease in participation of Black students. Yet this conclusion is incorrect; the growth in Hispanic CS degrees is likely also driven by the overall demographic shift in the country's college-going population.    Figure~\ref{HispanicBlackCSOutOfCohort} provides a more accurate picture of the extent to which each group is pursuing CS degrees.  In this figure we look at Hispanic CS degrees as a percentage of total Hispanic degrees and Black CS degrees as a percentage of total Black degrees, showing clearly that both groups began a steady increase in CS as a percentage of their cohort degrees as of 2010.


\begin{figure}
    \centering
    \includegraphics[width=3in]{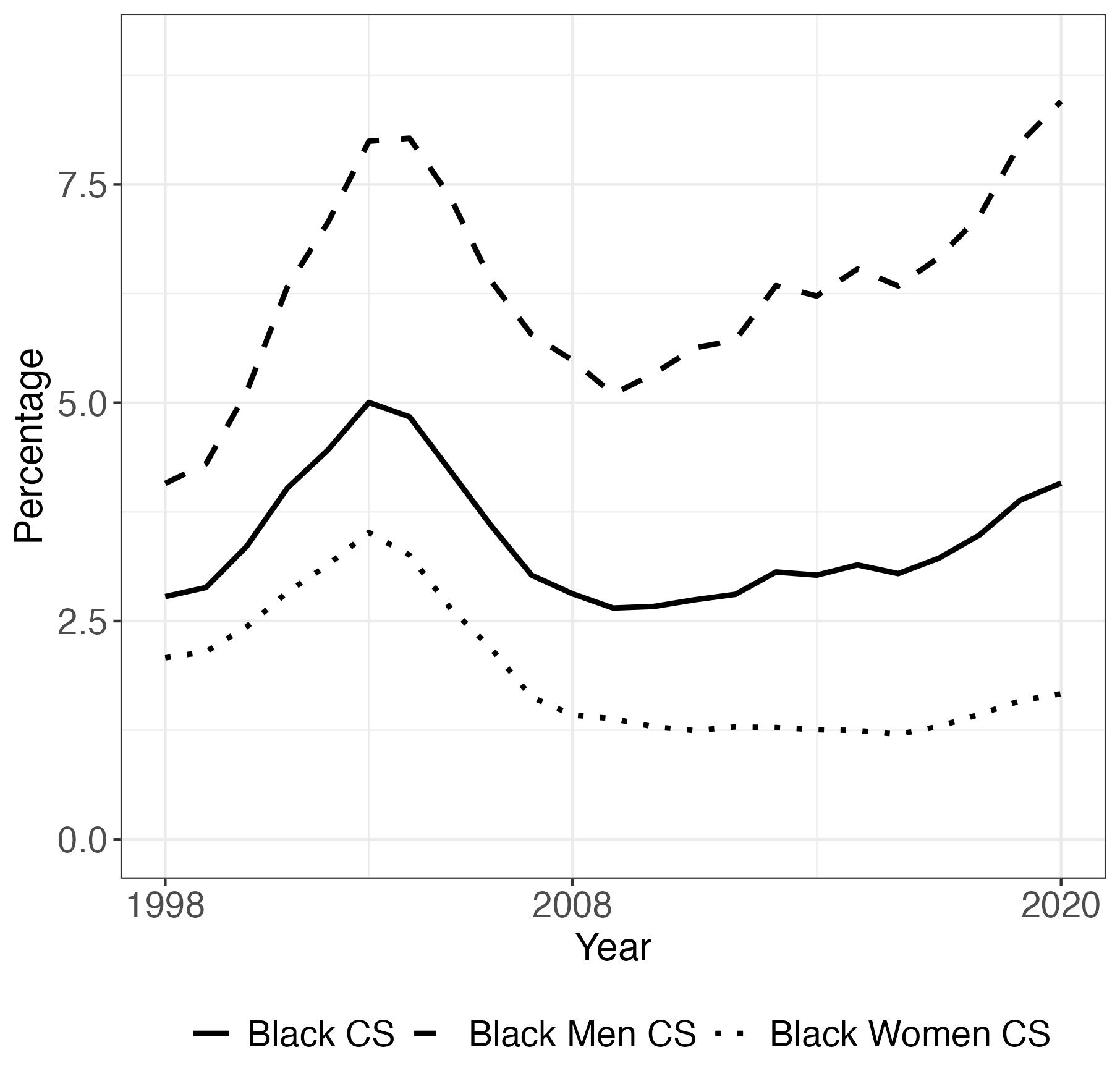}
    \caption{Black CS degrees, percentage of gender cohort.}
    \label{BlackCSOutOfGenderCohort}
\end{figure}

It is critically important that we look at data intersectionally. Figure~\ref{HispanicBlackCSOutOfCohort} shows an increase in CS degrees for Black students but does not address the question of whether that increase is reflected in both Black women's degrees and Black men's degrees.  Figure~\ref{BlackCSOutOfGenderCohort} reports Black CS degrees as percentage of all Black degrees, Black men's CS degrees as a percentage of all Black men's degrees, and Black women's CS degrees as a percentage of all Black women's degrees.  This clearly indicates that, while there is overall increase in the extent to which Black men are being attracted to and retained in CS, there is no analogous increase in the participation of Black women.  

These examples show that the standard analysis used to analyze degree data, namely examining a group's CS degrees as a percentage of all CS degrees, is faulty because the relative size of the component groups changes over time.  The standard analysis, therefore, can falsely indicate a negative trend that does not actually exist and, conversely, hide progress that is occurring.  Cohort analysis (across gender, race and ethnicity, and intersectional identities) gives an accurate picture of the extent to which groups are attracted into the field.  Interested readers can carry out intersectional cohort analysis for U.S. nationwide CIP-11 data or for any college or university via a webapp available at \url{https://aiice.shinyapps.io/AiiCE/}.  We next explore the importance of examining CS degree data in the context of university degree data.   

\section{The Importance of University Context}
\label{sec:context}

Computing departments often have no control over who attends their university, but they can influence who can {\em discover} computing, feel a sense of belonging, and persist to graduation.  Looking at the data intersectionally in their department in comparison to their university's overall data can let them see their ``opportunity gap''. 

Many departments struggle to gain access to the demographic data they need to track students by their intersectional identity as they make their way through the CS degree.\footnote{The Center for Inclusive Computing at Northeastern has supported 58 universities financially to unlock this centrally held data for all majors and enrollments in computing classes, illustrating that it is possible for CS departments to obtain this data.}   However, all departments have access to their graduation data via IPEDS \cite{IPEDS-2021}. To understand the importance of reporting intersectional data in the context of the university's data we first look at graduation data for the entire U.S., and then examine the opportunity gap for a Hispanic Serving Institution (HSI) in California.

\begin{figure}
    \centering
    \includegraphics[width=3.2in,height=2.5in]{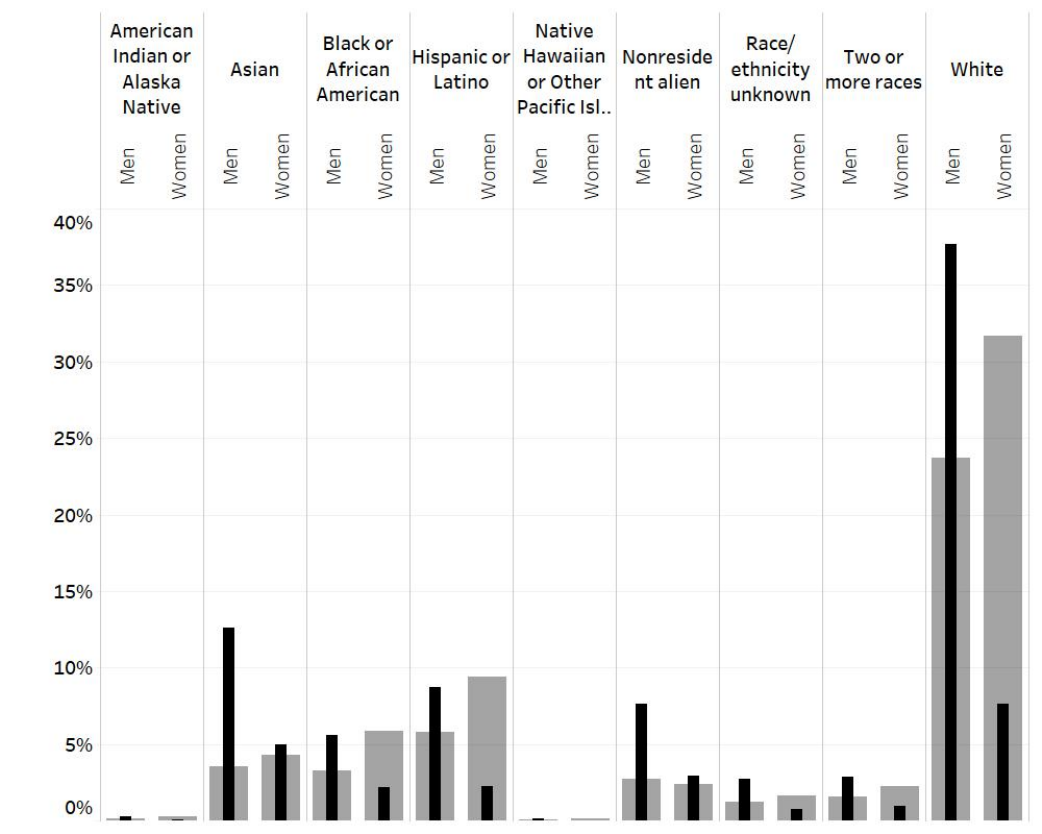}
    \caption{Intersectional 2021 IPEDS data of CS degrees (black solid bars) versus all degrees in the U.S. (gray bars).}
    \label{fig:IPEDS-2021}
\end{figure}

Figure \ref{fig:IPEDS-2021} shows the 2021 national computing graduation rates for the intersection of gender and race/ethnicity captured by IPEDS as the solid bar (black and in the foreground), and the graduation rates for all degrees as the shaded bar (gray and behind the solid bar).  For each race/ethnicity category tracked by IPEDS, the bar on the left represents men and the bar on the right represents women.  To understand this data, we focus on a particular intersectional identity. In 2021, 8\% of all computing graduates in the U.S. identified as Hispanic men, whereas only 2\% identified as Hispanic women.  In contrast, 6\% of graduates from university (in any field) identified as Hispanic men and 9\% as Hispanic Women.  Thus with respect to who graduated from university in the U.S. in 2021, Hispanic men are over-represented in computing (8\% versus 6\%) and Hispanic women are underrepresented (2\% versus 9\%).   

\begin{figure}
    \centering
    \includegraphics[width=3.8in,height=2.5in]{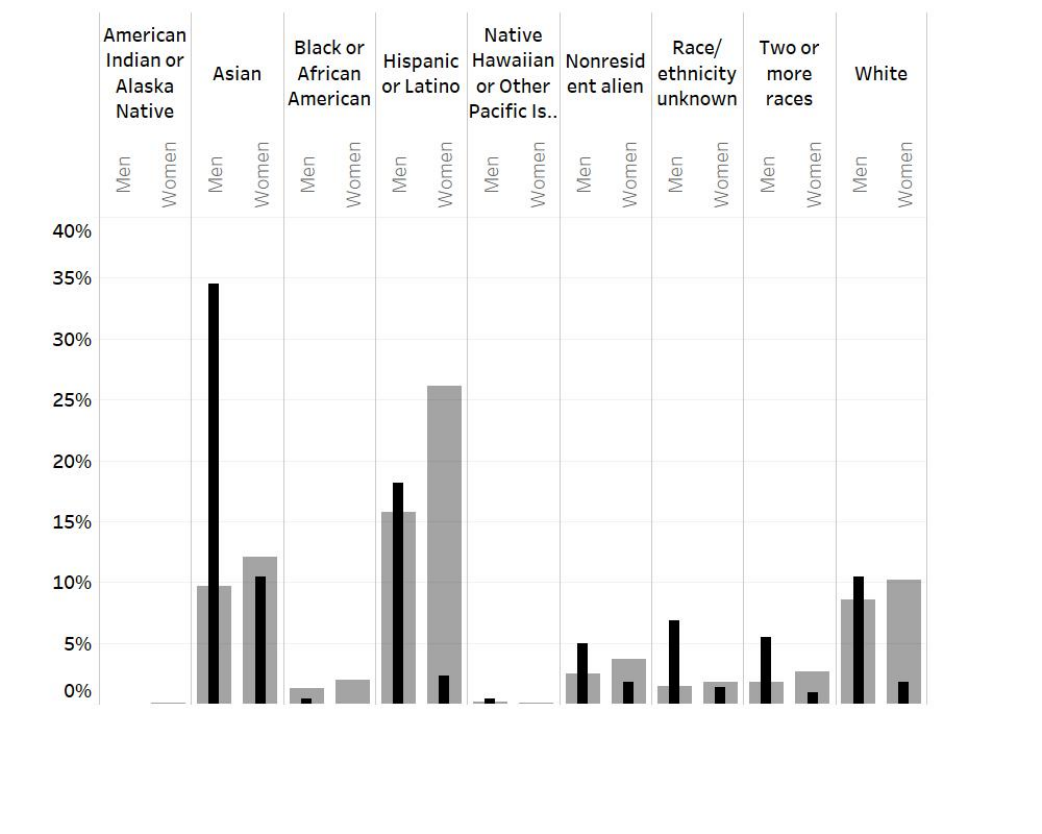}
    \caption{Intersectional IPEDS data of CS versus all degrees at an HSI in California.}
    \label{fig:HSI-2021}
\end{figure}

Computer science departments can understand their opportunity gap by looking at their own data in comparison to the data across their university.
In Figure \ref{fig:HSI-2021}, we show the 2021 graduation data for an HSI in California.  What is particularly striking is that out of the 2300+ Hispanic women that graduated from this university in 2021, {\em only 5 of them graduated with a CS degree}.  Indeed, this problem persists across women of all races and ethnicities at this university; in 2021, 59\% of its graduates were women, whereas only 19\% of CS graduates were women.

 \begin{table}
     \centering
     \begin{tabular}{p{1.5in}|c|c}
         & Institution 1 (HSI) & Institution 2\\ \hline
         Total CS degrees & 112 & 580 \\ \hline
         Hispanic \%-age of CS degrees & 90.2 & 3.9 \\ \hline
         Hispanic women \%-age of CS degrees & 14.3 & 1.7 \\ \hline
              \end{tabular}
     \caption{Comparison of two universities - standard analysis.}
     \label{TwoUnivCompareNoCohort}
 \end{table}

Comparing diversity in computing across different institutions is best done in the context of their university demographics.  In Table ~\ref{TwoUnivCompareNoCohort} we use the standard analysis to compare two institutions, both public universities.  Institution 1 is an HSI  whereas Institution 2 is not dominated by any single race or ethnicity grouping.  Unsurprising at Institution 1, since they are the majority of the student body, Hispanic students earn the majority of CS degrees.  Similarly, because the Hispanic student body is very large at Institution 1, we can see that Hispanic women make up a much larger proportion of total CS degrees than they do at Institution 2.  Yet when we apply a cohort analysis approach (from Section \ref{sec:cohort}) we generate a picture of these two institutions that points more clearly to where existing interventions may be effective and what new interventions might be useful.  As we can see in Table~\ref{TwoUnivCompareWithCohort}, the percentage of Hispanic student degrees that are earned in CS is similar at both institutions (3.2\% versus 3.4\%).  Yet interesting differences arise when we consider the intersectionality of gender with ethnicity.  We can see that at Institution 1, 0.9\% of all Hispanic women's degrees are earned in CS whereas in Institution 2,  2.7\% of all Hispanic women's degrees are earned in CS.  In contrast, Institution 1 does a better job drawing Hispanic men into CS (6.4\% of Hispanic men's degrees) compared to Institution 2 (4.3\% of Hispanic men's degrees).  This may indicate that Institution 2 has strong interventions designed to recruit and retain women in computing, with derivative impact on Hispanic women students, but does not necessarily have efforts targeting students from historically marginalized race and ethnicity groups.  By the same token, it would appear that Institution 1 should consider developing interventions focused on their women students.

\begin{table}
     \centering
     \begin{tabular}{p{1.5in}|c|c}
         & Institution 1 (HSI) & Institution 2\\ \hline
         Hispanic CS as \%-age of all Hispanic degrees & 3.2 & 3.4 \\ \hline
         Hispanic women's CS as \%-age of all Hispanic women's degrees & 0.9 & 2.7 \\ \hline
         Hispanic men's CS as \%-age of all Hispanic men's degrees & 6.4 & 4.3. \\ \hline
              \end{tabular}
     \caption{Comparison of two universities - cohort analysis.}
     \label{TwoUnivCompareWithCohort}
 \end{table}

The examples in this section illustrate the utility of looking at intersectional graduation data of the CS department in the context of the overall demographics of the university. This provides them with the opportunity gap they can tackle.  For example, a CS department that awards 30\% of degrees to women is a stunning success in a technical university  where the representation of women across {\em all} degrees is 30\% but represents an opportunity gap in a university that is 57\% women.\footnote{Clearly the technical university should work on their overall representation of women across all degrees, but this is often outside of the influence of the CS department.} In the next section we examine the utility of summary statistics for measuring diversity using entropy-based measures/metrics.

\section{Entropy-Based Diversity Metrics}

There is a large body of literature on the use of entropy-based metrics for measuring diversity in populations.  Perhaps the most analogous application of these metrics to BPC is the work done on measuring residential segregation. Massey and Denton, for example, discuss many metrics used to measure residential segregation \cite{Massey-Denton-88}. They proposed that residential segregation could be described by five dimensions: evenness, exposure, concentration, centralization, and clustering. 
Their analysis has been replicated and widely discussed \cite{Massey-Denton-88, Massey-White-Phua-1996, Massey-2012,Cortese-1976, White-1983, White-1986}, and serves as a good baseline for measuring diversity in computing programs.\footnote{See \cite{White-1986} for a discussion of the desirable properties of a diversity index, as originally stated by Pielou \cite{Pielou-1975} and later expanded by White \cite{White-1986}.}

One of these dimensions, evenness, has a good parallel with our analysis of diversity in computing. Recently Kelly lamented that there was “no composite institution-level measure for ethnic diversity”~\cite[p.~41]{Kelly-2019} and proposed to use the Shannon index as a way to measure ethnic distribution in academic programs. Kelly goes on to state “What is needed is a single index that does more than simply count how many ethnicities exist in a dataset, but instead takes account of the relative population size of those different ethnicities”~\cite[p.~42]{Kelly-2019}.








The Shannon information index \cite{White-1986}, or the Entropy index, has been commonly used for such purpose. This measure is defined as:

 \begin{center}
     $H = - \sum_{k=1}^{k}{p_{k}~*~ln~p_{k}}$ 
 \end{center}

\vspace{0.1in}

\noindent where $k$ is the number of groups in the analysis, $N_{k}$ is the number of students in group $k$, $N$ is the total number of students in the population, and $p_{k}  = N_{k} / N$ (i.e., the percentage of group $k$ in the population).  When all $k$ groups are equally distributed $H$ is maximized. For our purposes we will use the normalized version, called the Shannon Equitability Index (see \cite{Kelly-2019}) which is computed as $E_H = H/ln(k)$ and produces values between 0 (no diversity) and 1 (all groups are equally represented). $E_H$ (often called “Evenness”) represents the degree to which all groups are equally proportioned in the population of study. When this value is represented as a percent (0\% .. 100\%) it can be interpreted as the percent of a uniform distribution that a particular distribution represents.



\begin{figure}
    \centering
    \includegraphics[width=3.2in]{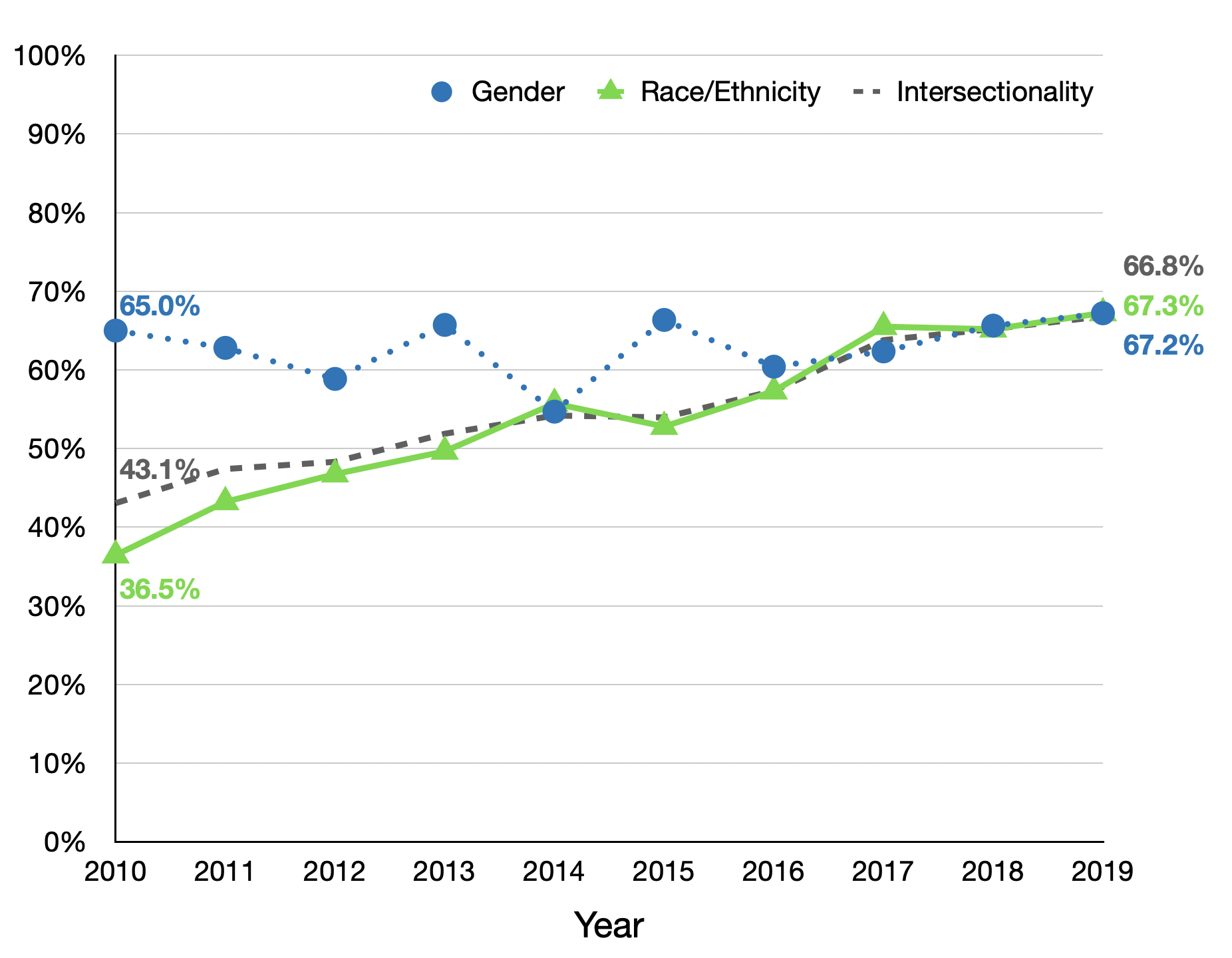}
    \caption{$E_H$ for gender, race, and intersectionality for 2010-2019 graduation data for Univ-11 (see Table~\ref{Top12NCUniversities} for designation)}
    \label{fig:CIPEquitability}
\end{figure}

To illustrate the strengths and weaknesses of $E_H$ for our purposes, we apply this measure to the IPEDS data of the 12 universities in North Carolina that graduated the largest number of students with CS degrees in 2020. These institutions are listed in Table~\ref{Top12NCUniversities}  and include public and private, urban and rural, Historically Black Colleges and Universities (HBCUs), as well as different levels in the Carnegie Classification.\footnote{ https://carnegieclassifications.acenet.edu.}
\begin{table}
     \centering
     \begin{tabular}{l|p{1.75in}|l}
        Univ.&Carnegie Classification&Additional Info\\ \hline \hline
        Univ-1&Master's - Larger Programs&Public, Rural\\ \hline
        Univ-2&Baccalaureate - Arts \& Sciences&Private\\ \hline
        Univ-3&Doctorate - Very High Research&Private\\ \hline
        Univ-4&Doctorate - Research&Public\\ \hline
        Univ-5&Baccalaureate - Arts \& Sciences&Private, HBCU\\ \hline
        Univ-6&Doctorate - High Research&Public, HBCU\\ \hline
        Univ-7&Master's - Larger Programs&Public, HBCU\\ \hline
        Univ-8&Doctorate - Very High Research&Public\\ \hline
        Univ-9&Baccalaureate - Arts \& Sciences&Public\\ \hline
        Univ-10&Doctorate - Very High Research&Public\\ \hline
        Univ-11&Doctorate - Research&Public\\ \hline
        Univ-12&Doctorate - High Research&Public\\ \hline
      \end{tabular}
     \caption{Twelve largest universities awarding CS degrees in North Carolina}
     \label{Top12NCUniversities}
 \end{table}
Figure \ref{fig:CIPEquitability} 
shows $E_H$ for the CS program at Univ-11 for the years 2010-2019. It shows three different calculations of $E_H$: 1)  gender (male/female); 2) race/ethnicity (American Indian or Alaska Native, Asian, Black or African American, Hispanic or Latino, Native Hawaiian or Other Pacific Islander, White, and two or more races); and 3) intersectionality which includes all combinations of gender and race. For this institution, you can see that $E_H$ calculated for gender has not changed much;  rising to 67.2\% in 2019 from  65\% in 2010. In contrast when we look at race, we observe that in the same time period $E_H$ has risen from 36.5\% in 2010 to 67.3\% in 2019, which is a significant improvement. This example shows that $E_H$ can be used to track diversity in a single institution over time.  The next analysis illustrates the weaknesses of $E_H$ as a measure to compare diversity {\em across} institutions.


\begin{figure}
    \centering
    \includegraphics[width=3.2in]{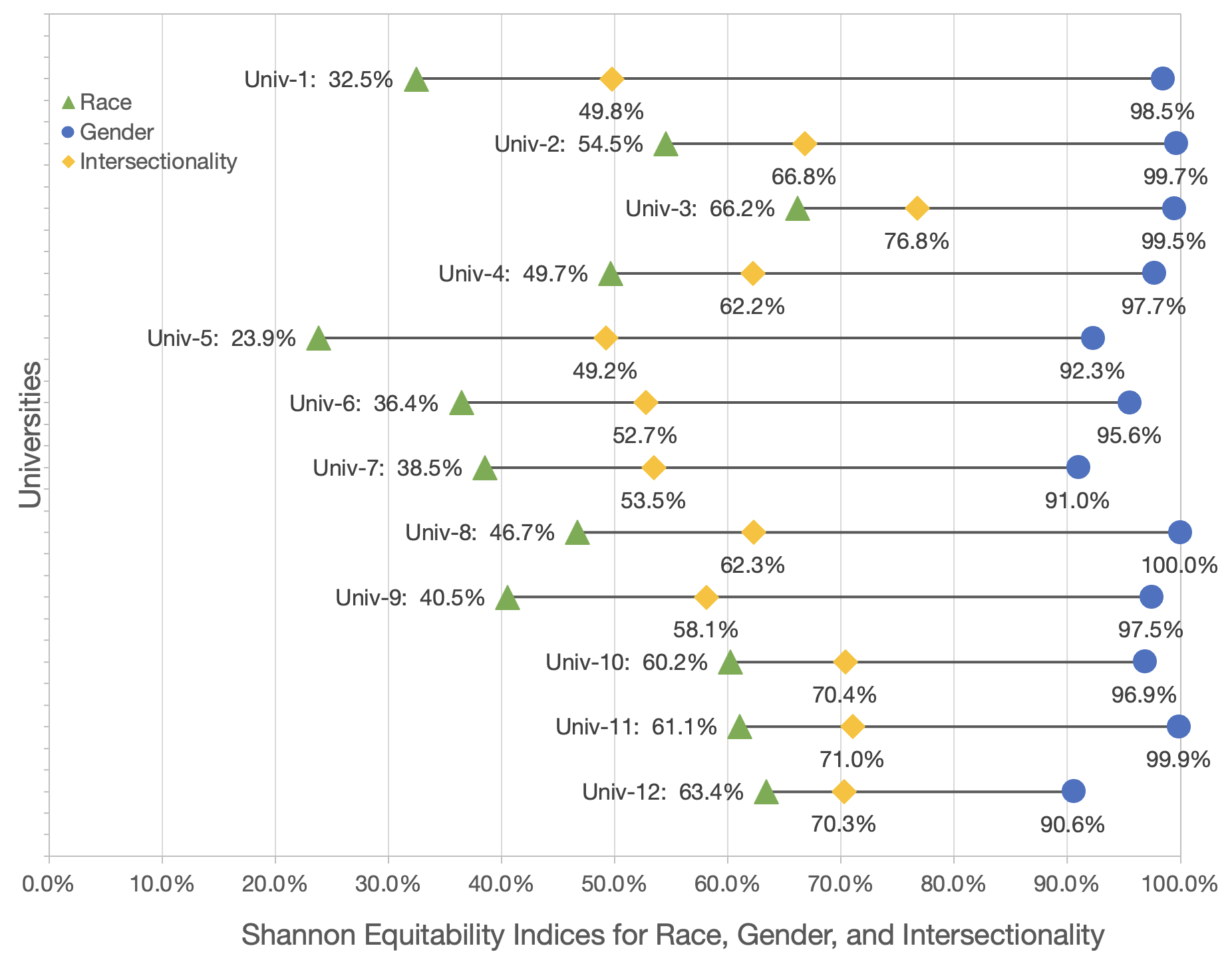}
    \caption{$E_H$ for the graduation data for the institutions listed in Table~\ref{Top12NCUniversities} examining race (triangle), gender (circle), and intersectionality (diamond).}
    \label{fig:NCUniversities}
\end{figure}

Figure \ref{fig:NCUniversities} shows a dumbbell graph of $E_H$ for all institutions shown in Table~\ref{Top12NCUniversities} for gender (circle), race/ethnicity (triangle) and the intersection of race and gender (diamond) of the 2020 graduation data across {\em all degrees} for each university as reported in IPEDS.   As shown in the figure, nearly all institutions are close to gender parity with 90.6\% as the lowest $E_H$ value among this group. Indeed for these universities, the percentage of female graduates in 2020 ranges from 49\% to 68\% of all students on campus. Note that a student body which is 50\% female and 50\% male would yield 100\% for the $E_H$ metric.

The $E_H$ metric for race tells a different story. $E_H$ calculated using race ranges from 23\% to 66\% for this set of institutions. It is worth noting that the three of the institutions with lowest value for the race $E_H$ metric are all HBCUs where the representation of African-American students ranges between 81\% and 93\%. $E_H$ measures how close a population is to an uniform distribution, ignoring the context of the institution. Indeed, its is expected that HBCUs would have a low value in the $E_H$ metric given the mission and composition of HBCUs.  Therefore, we must be careful when when using $E_H$ to compare {\em across} institutions because it ignores institutional context.

As we saw in Sections \ref{sec:cohort} and \ref{sec:context}, to truly understand participation of a particular group within a particular computing program we must consider the representation of sub-populations in the context of the larger reference group (i.e., cohort analysis).  $E_H$ as a measure of evenness ignores the size of the reference group. For that, we turn to the Jensen-Shannon divergence \cite{Lin-1991}, which measures the similarity between two probability distributions. It is based on the Kullback–Leibler divergence \cite{KL}, with some notable (and useful) differences, including that it is symmetric and it always has a finite value. The square root of the Jensen–Shannon divergence is a metric often referred to as Jensen–Shannon (JS) distance.\footnote{To compare two probability distributions $P$ and $Q$, the Jensen-Shannon divergence, JSD, is computed as: $JSD = H(M) - \frac{1}{2} (H(P)) + H(Q))$, where $M = \frac{1}{2}(P+Q)$ and $H(P)$ is the Shannon Entropy as defined above.  For a more detailed explanation please see https://en.wikipedia.org/wiki/Jensen-Shannon\_divergence.}

\begin{figure}
    \centering
    \includegraphics[width=3.2in]{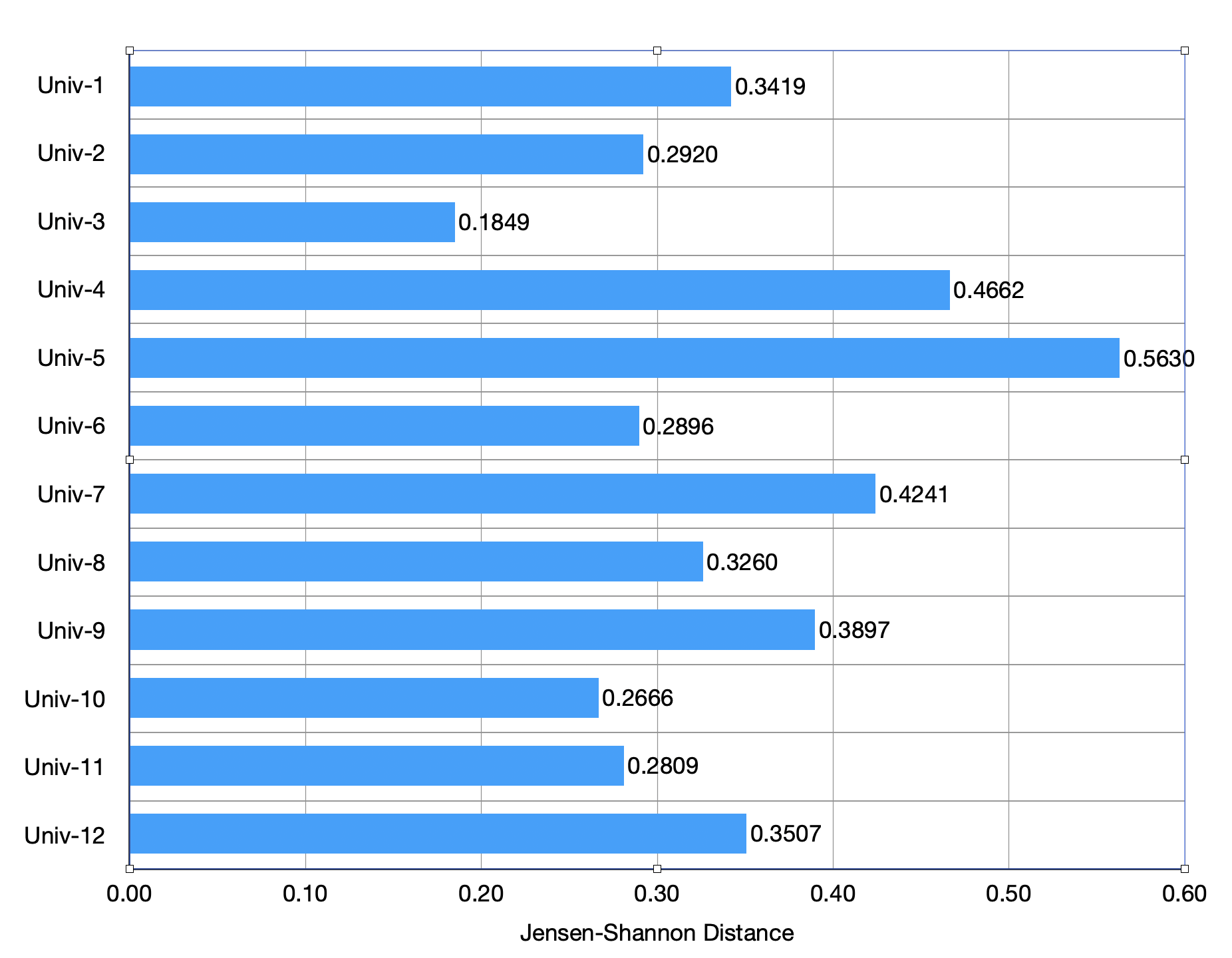}
    \caption{Jensen-Shannon distance between the distribution of CIP 11 degrees awarded when compared with the distribution of all degrees awarded at 12 institutions in North Carolina. Lower values indicate smaller distances between the two distributions.}
    \label{fig:JSDistance}
\end{figure}

\begin{figure}
    \centering
    \includegraphics[width=3.2in]{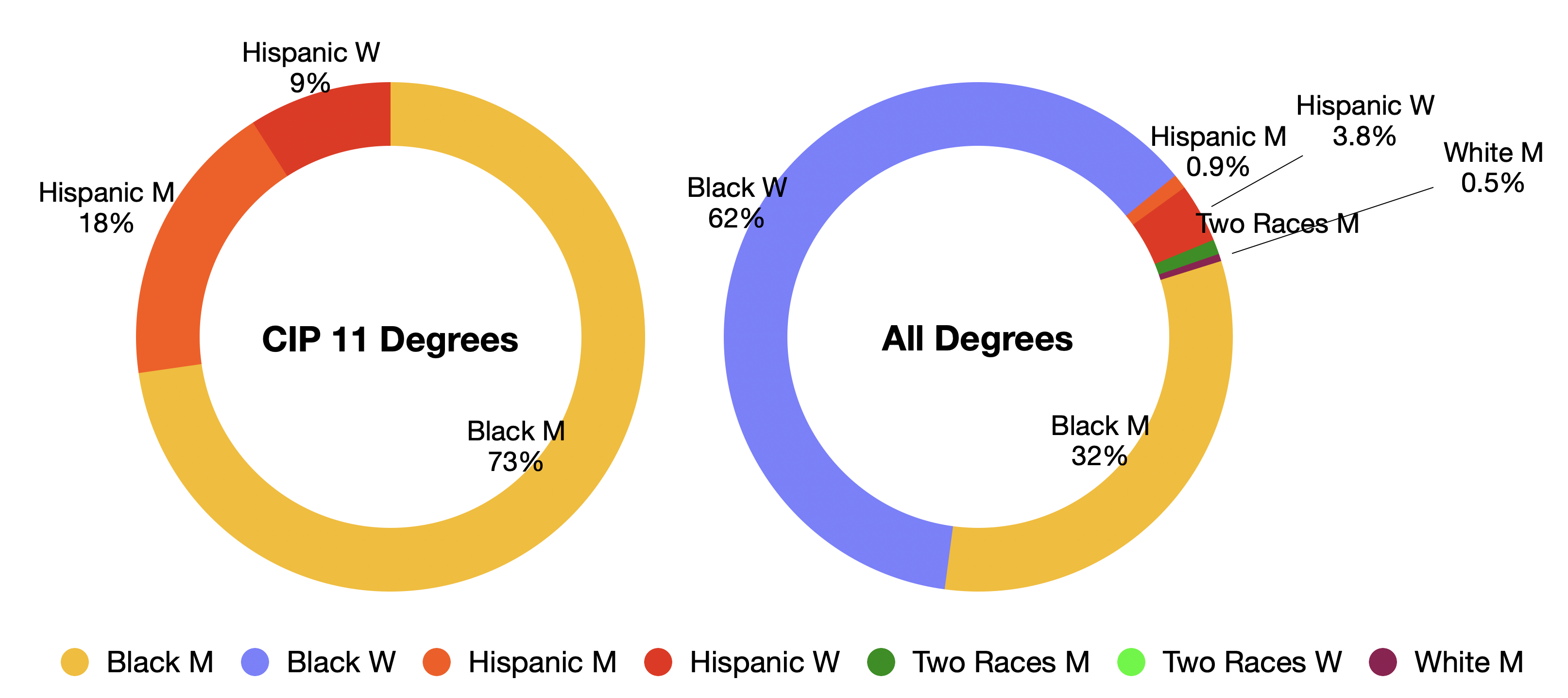}
    \caption{Intersectional distribution of graduates in CIP 11 (left chart) and of graduates from all degrees (right chart) at Univ-5.}
    \label{fig:DistributionDifferences}
\end{figure}

Figure \ref{fig:JSDistance} shows the JS distance between the intersectional distribution of CIP 11 degrees awarded and all degrees awarded for each of the institutions listed in Table~\ref{Top12NCUniversities}. A value of zero means that the two distributions are identical.
To understand this metric, we look at Univ-5, a private HBCU, which has the highest JS distance of the 12 universities. Figure~\ref{fig:DistributionDifferences} shows the intersectional breakdown for CIP 11 degrees and all degrees for Univ-5. Although the majority of degrees awarded (62\%) by this institution went to Black women, the institution awarded {\em zero} CIP 11 degrees to Black women. Furthermore, Hispanic men and women are overrepresented in CIP 11 w.r.t. to all degrees awarded on this campus.

In this section we examined the use of two commonly applied entropy-based measures for evaluating the demographic diversity of a population.  The first, $E_H$, is maximized when all sub-populations are uniformly distributed.  The second, the JS distance, measures how different the population of CS is compared to the reference population and can be seen as a summary statistic of the data presented in Figure~\ref{fig:IPEDS-2021}.  Both are useful summary statistics to track over time to see if representation is increasing overall, but are best used in combination with the other more detailed analysis methods presented.    

\section{Conclusion}

In this paper we have pointed out the limitations of looking at diversity and assessing BPC efforts via the single metric of the percentage of each sub-population's degree attainment as a proportion of the total degrees in the field.  We make three recommendations for quantitative data analysis of BPC efforts.  First, we need to examine cohort-based data to evaluate each group's interest in computing, independent of larger demographic shifts in the student population.  Second, the field as a whole needs to adopt the norm of always reporting intersectional data, rather than just looking at men/women and race/ethnicity separately.  Third, university demographic context must be considered when evaluating how well a computing department is doing to broaden participation, thereby also accounting for shifts in the overall university population.    Cohort-based analysis, intersectionality, and entropy based measures provide different insights that are necessary to fully understand the challenges and successes of BPC activities. 

We conclude with one final observation about what data to analyze.  In this paper we analyzed IPEDS graduation data, but there are additional facets to the challenges faced by students not captured in this data. These include: differential exposure to computing, recruitment, mentoring, retention, and institutional barriers students face in discovering and majoring in computing \cite{Brodley-RESPECT-2021}.  Graduation data is not sufficient for monitoring the impact of BPC activities, particularly on the introductory sequence of courses in the computing major. Thus we recommend tracking the intersectional demographics of the drop/fail/withdraw rates of students in the  introductory sequence classes by professor  {\em every semester/quarter} to uncover opportunities for change in the curriculum, the co-curricular elements, etc. We note that it can be difficult for computing departments to obtain the often centrally held student demographic data.  The Center for Inclusive Computing has, to date, successfully helped 58 U.S. universities provide this intersectional data for their computing departments.    Finally, we would be remiss in not pointing out that, in addition to quantitative analysis, we must also examine qualitative analysis and student survey data to better understand the experiences of students and the opportunities for our programs to be more inclusive.


\begin{acks}
This material is based in part upon work supported by the National Science Foundation under Grant Nos. 2216629 and 2118453, in part based upon work done by the third author while serving at the National Science Foundation. We would like to thank Jodi Tims and Christine Alvarado for their close read of an earlier version of this paper.
\end{acks}

\bibliographystyle{ACM-Reference-Format}
\bibliography{bib}


\begin{thebibliography}{20}


\ifx \showCODEN    \undefined \def \showCODEN     #1{\unskip}     \fi
\ifx \showDOI      \undefined \def \showDOI       #1{#1}\fi
\ifx \showISBNx    \undefined \def \showISBNx     #1{\unskip}     \fi
\ifx \showISBNxiii \undefined \def \showISBNxiii  #1{\unskip}     \fi
\ifx \showISSN     \undefined \def \showISSN      #1{\unskip}     \fi
\ifx \showLCCN     \undefined \def \showLCCN      #1{\unskip}     \fi
\ifx \shownote     \undefined \def \shownote      #1{#1}          \fi
\ifx \showarticletitle \undefined \def \showarticletitle #1{#1}   \fi
\ifx \showURL      \undefined \def \showURL       {\relax}        \fi
\providecommand\bibfield[2]{#2}
\providecommand\bibinfo[2]{#2}
\providecommand\natexlab[1]{#1}
\providecommand\showeprint[2][]{arXiv:#2}

\bibitem[Barr(2018)]%
        {Barr-Inroads-2018}
\bibfield{author}{\bibinfo{person}{Valerie Barr}.} \bibinfo{year}{2018}\natexlab{}.
\newblock \showarticletitle{Different Denominators, Different Results: Reanalyzing CS Degrees by Gender, Race, and Ethnicity}.
\newblock \bibinfo{journal}{\emph{ACM Inroads}} \bibinfo{volume}{9}, \bibinfo{number}{3} (\bibinfo{date}{September} \bibinfo{year}{2018}), \bibinfo{pages}{40--47}.
\newblock
\showISSN{2153-2192}
\urldef\tempurl%
\url{https://doi.org/10.1145/2891414}
\showDOI{\tempurl}


\bibitem[Brodley et~al\mbox{.}(2021)]%
        {Brodley-RESPECT-2021}
\bibfield{author}{\bibinfo{person}{Carla~E. Brodley}, \bibinfo{person}{Catherine Gill}, {and} \bibinfo{person}{Sally Wynn}.} \bibinfo{year}{2021}\natexlab{}.
\newblock \showarticletitle{Diagnosing why Representation Remains Elusive at your University: Lessons Learned from the Center for Inclusive Computing{\textquotesingle}s Site Visits}. In \bibinfo{booktitle}{\emph{2021 Conference on Research in Equitable and Sustained Participation in Engineering, Computing, and Technology ({RESPECT})}}. \bibinfo{publisher}{{IEEE}}, \bibinfo{address}{Philadelphia, PA, USA}, \bibinfo{pages}{1--4}.
\newblock
\urldef\tempurl%
\url{https://doi.org/10.1109/respect51740.2021.9620552}
\showDOI{\tempurl}


\bibitem[Cortese et~al\mbox{.}(1976)]%
        {Cortese-1976}
\bibfield{author}{\bibinfo{person}{Charles~F. Cortese}, \bibinfo{person}{R.~Frank Falk}, {and} \bibinfo{person}{Jack~K. Cohen}.} \bibinfo{year}{1976}\natexlab{}.
\newblock \showarticletitle{Further Considerations on the Methodological Analysis of Segregation Indices}.
\newblock \bibinfo{journal}{\emph{American Sociological Review}} \bibinfo{volume}{41}, \bibinfo{number}{4} (\bibinfo{year}{1976}), \bibinfo{pages}{630--637}.
\newblock
\showISSN{00031224}
\urldef\tempurl%
\url{http://www.jstor.org/stable/2094840}
\showURL{%
\tempurl}


\bibitem[Crenshaw(1989)]%
        {Crenshaw1989}
\bibfield{author}{\bibinfo{person}{Kimberlé~W. Crenshaw}.} \bibinfo{year}{1989}\natexlab{}.
\newblock \showarticletitle{Demarginalizing the Intersection of Race and Sex: A Black Feminist Critique of Antidiscrimination Doctrine, Feminist Theory and Antiracist Politics}.
\newblock \bibinfo{journal}{\emph{University of Chicago Legal Forum}} \bibinfo{volume}{1989}, \bibinfo{number}{1} (\bibinfo{year}{1989}), \bibinfo{pages}{139--167}.
\newblock


\bibitem[Kelly(2019)]%
        {Kelly-2019}
\bibfield{author}{\bibinfo{person}{Anthony Kelly}.} \bibinfo{year}{2019}\natexlab{}.
\newblock \showarticletitle{A new composite measure of ethnic diversity: Investigating the controversy over minority ethnic recruitment at Oxford and Cambridge universities}.
\newblock \bibinfo{journal}{\emph{British Educational Research Journal}} \bibinfo{volume}{45}, \bibinfo{number}{1} (\bibinfo{year}{2019}), \bibinfo{pages}{41--82}.
\newblock
\urldef\tempurl%
\url{https://doi.org/10.1002/berj.3482}
\showURL{%
\tempurl}


\bibitem[Kullback and Leibler(1951)]%
        {KL}
\bibfield{author}{\bibinfo{person}{S. Kullback} {and} \bibinfo{person}{R.A. Leibler}.} \bibinfo{year}{1951}\natexlab{}.
\newblock \showarticletitle{On information and sufficiency}.
\newblock \bibinfo{journal}{\emph{Annals of Mathematical Statistics}} \bibinfo{volume}{22}, \bibinfo{number}{1} (\bibinfo{year}{1951}), \bibinfo{pages}{79--86}.
\newblock


\bibitem[Lin(1991)]%
        {Lin-1991}
\bibfield{author}{\bibinfo{person}{J. Lin}.} \bibinfo{year}{1991}\natexlab{}.
\newblock \showarticletitle{Divergence measures based on the Shannon entropy}.
\newblock \bibinfo{journal}{\emph{IEEE Trans. Information Theory}} \bibinfo{volume}{37}, \bibinfo{number}{1} (\bibinfo{year}{1991}), \bibinfo{pages}{145--151}.
\newblock


\bibitem[Massey(2012)]%
        {Massey-2012}
\bibfield{author}{\bibinfo{person}{Douglas~S. Massey}.} \bibinfo{year}{2012}\natexlab{}.
\newblock \showarticletitle{Reflections on the Dimensions of Segregation}.
\newblock \bibinfo{journal}{\emph{Social forces; a scientific medium of social study and interpretation}} \bibinfo{volume}{91}, \bibinfo{number}{1} (\bibinfo{year}{2012}), \bibinfo{pages}{39--43}.
\newblock
\urldef\tempurl%
\url{https://doi.org/10.1093/sf/sos118}
\showDOI{\tempurl}


\bibitem[Massey and Denton(1988)]%
        {Massey-Denton-88}
\bibfield{author}{\bibinfo{person}{Douglas~S. Massey} {and} \bibinfo{person}{Nancy~A. Denton}.} \bibinfo{year}{1988}\natexlab{}.
\newblock \showarticletitle{The Dimensions of Residential Segregation}.
\newblock \bibinfo{journal}{\emph{Social Forces}} \bibinfo{volume}{67}, \bibinfo{number}{2} (\bibinfo{year}{1988}), \bibinfo{pages}{281--315}.
\newblock
\showISSN{00377732, 15347605}
\urldef\tempurl%
\url{http://www.jstor.org/stable/2579183}
\showURL{%
\tempurl}


\bibitem[Massey et~al\mbox{.}(1996)]%
        {Massey-White-Phua-1996}
\bibfield{author}{\bibinfo{person}{Douglas~S. Massey}, \bibinfo{person}{Michael~J. White}, {and} \bibinfo{person}{Voon chin Phua}.} \bibinfo{year}{1996}\natexlab{}.
\newblock \showarticletitle{The Dimensions of Segregation Revisited}.
\newblock \bibinfo{journal}{\emph{Sociological Methods \& Research}} \bibinfo{volume}{25}, \bibinfo{number}{2} (\bibinfo{year}{1996}), \bibinfo{pages}{172--206}.
\newblock
\urldef\tempurl%
\url{https://doi.org/10.1177/0049124196025002002}
\showDOI{\tempurl}


\bibitem[Ovalle et~al\mbox{.}(2023)]%
        {Ovalle2023}
\bibfield{author}{\bibinfo{person}{Anaelia Ovalle}, \bibinfo{person}{Arjun Subramonian}, \bibinfo{person}{Vagrant Gautam}, \bibinfo{person}{Gilbert Gee}, {and} \bibinfo{person}{Kai-Wei Chang}.} \bibinfo{year}{2023}\natexlab{}.
\newblock \showarticletitle{Factoring the Matrix of Domination: A Critical Review and Reimagination of Intersectionality in AI Fairness}. In \bibinfo{booktitle}{\emph{Proceedings of the 2023 AAAI/ACM Conference on AI, Ethics, and Society}} (Montr\'{e}al, QC, Canada) \emph{(\bibinfo{series}{AIES '23})}. \bibinfo{publisher}{Association for Computing Machinery}, \bibinfo{address}{New York, NY, USA}, \bibinfo{pages}{496–511}.
\newblock
\showISBNx{9798400702310}
\urldef\tempurl%
\url{https://doi.org/10.1145/3600211.3604705}
\showDOI{\tempurl}


\bibitem[Pielou(1975)]%
        {Pielou-1975}
\bibfield{author}{\bibinfo{person}{E.~C. Pielou}.} \bibinfo{year}{1975}\natexlab{}.
\newblock \bibinfo{booktitle}{\emph{Ecological diversity}}.
\newblock \bibinfo{publisher}{John Wiley}, \bibinfo{address}{New York, New York, USA.}
\newblock


\bibitem[Rankin and Thomas(2019)]%
        {Rankin2019}
\bibfield{author}{\bibinfo{person}{Yolanda~A. Rankin} {and} \bibinfo{person}{Jakita~O. Thomas}.} \bibinfo{year}{2019}\natexlab{}.
\newblock \showarticletitle{Straighten up and Fly Right: Rethinking Intersectionality in HCI Research}.
\newblock \bibinfo{journal}{\emph{Interactions}} \bibinfo{volume}{26}, \bibinfo{number}{6} (\bibinfo{date}{oct} \bibinfo{year}{2019}), \bibinfo{pages}{64–68}.
\newblock
\showISSN{1072-5520}
\urldef\tempurl%
\url{https://doi.org/10.1145/3363033}
\showDOI{\tempurl}


\bibitem[Rankin et~al\mbox{.}(2021)]%
        {Rankin2021}
\bibfield{author}{\bibinfo{person}{Yolanda~A. Rankin}, \bibinfo{person}{Jakita~O. Thomas}, {and} \bibinfo{person}{Sheena Erete}.} \bibinfo{year}{2021}\natexlab{}.
\newblock \showarticletitle{Black Women Speak: Examining Power, Privilege, and Identity in CS Education}.
\newblock \bibinfo{journal}{\emph{ACM Trans. Comput. Educ.}} \bibinfo{volume}{21}, \bibinfo{number}{4}, Article \bibinfo{articleno}{26} (\bibinfo{date}{oct} \bibinfo{year}{2021}), \bibinfo{numpages}{31}~pages.
\newblock
\urldef\tempurl%
\url{https://doi.org/10.1145/3451344}
\showDOI{\tempurl}


\bibitem[Thomas et~al\mbox{.}(2018)]%
        {Thomas2018}
\bibfield{author}{\bibinfo{person}{Jakita~O. Thomas}, \bibinfo{person}{Nicole Joseph}, \bibinfo{person}{Arian Williams}, \bibinfo{person}{Chan’tel Crum}, {and} \bibinfo{person}{Jamika Burge}.} \bibinfo{year}{2018}\natexlab{}.
\newblock \showarticletitle{Speaking Truth to Power: Exploring the Intersectional Experiences of Black Women in Computing}. In \bibinfo{booktitle}{\emph{2018 Research on Equity and Sustained Participation in Engineering, Computing, and Technology (RESPECT)}}. \bibinfo{publisher}{{IEEE}}, \bibinfo{address}{Baltimore, MD, USA}, \bibinfo{pages}{1--8}.
\newblock
\urldef\tempurl%
\url{https://doi.org/10.1109/RESPECT.2018.8491718}
\showDOI{\tempurl}


\bibitem[Trauth et~al\mbox{.}(2012)]%
        {Trauth2012}
\bibfield{author}{\bibinfo{person}{Eileen~M. Trauth}, \bibinfo{person}{Curtis Cain}, \bibinfo{person}{K.~D. Joshi}, \bibinfo{person}{Lynette Kvasny}, {and} \bibinfo{person}{Kayla Booth}.} \bibinfo{year}{2012}\natexlab{}.
\newblock \showarticletitle{Understanding Underrepresentation in IT through Intersectionality}. In \bibinfo{booktitle}{\emph{Proceedings of the 2012 IConference}} (Toronto, Ontario, Canada) \emph{(\bibinfo{series}{iConference '12})}. \bibinfo{publisher}{Association for Computing Machinery}, \bibinfo{address}{New York, NY, USA}, \bibinfo{pages}{56–62}.
\newblock
\showISBNx{9781450307826}
\urldef\tempurl%
\url{https://doi.org/10.1145/2132176.2132184}
\showDOI{\tempurl}


\bibitem[U.S. Department~of Education(2021)]%
        {IPEDS-2021}
\bibfield{author}{\bibinfo{person}{National Center for Education~Statistics U.S. Department~of Education}.} \bibinfo{year}{2021}\natexlab{}.
\newblock \bibinfo{booktitle}{\emph{The Integrated Postsecondary Education Data System}}.
\newblock \bibinfo{publisher}{Department of Education}, \bibinfo{address}{Washington, DC}.
\newblock
\urldef\tempurl%
\url{https://nces.ed.gov/ipeds/}
\showURL{%
\tempurl}


\bibitem[Warner et~al\mbox{.}(2021)]%
        {Warner2021}
\bibfield{author}{\bibinfo{person}{Jayce~R. Warner}, \bibinfo{person}{Joshua Childs}, \bibinfo{person}{Carol~L. Fletcher}, \bibinfo{person}{Nicole~D. Martin}, {and} \bibinfo{person}{Michelle Kennedy}.} \bibinfo{year}{2021}\natexlab{}.
\newblock \showarticletitle{Quantifying Disparities in Computing Education: Access, Participation, and Intersectionality}. In \bibinfo{booktitle}{\emph{Proceedings of the 52nd ACM Technical Symposium on Computer Science Education}} (Virtual Event, USA) \emph{(\bibinfo{series}{SIGCSE '21})}. \bibinfo{publisher}{Association for Computing Machinery}, \bibinfo{address}{New York, NY, USA}, \bibinfo{pages}{619–625}.
\newblock
\showISBNx{9781450380621}
\urldef\tempurl%
\url{https://doi.org/10.1145/3408877.3432392}
\showDOI{\tempurl}


\bibitem[White(1983)]%
        {White-1983}
\bibfield{author}{\bibinfo{person}{Michael~J. White}.} \bibinfo{year}{1983}\natexlab{}.
\newblock \showarticletitle{The Measurement of Spatial Segregation}.
\newblock \bibinfo{journal}{\emph{Amer. J. Sociology}} \bibinfo{volume}{88}, \bibinfo{number}{5} (\bibinfo{year}{1983}), \bibinfo{pages}{1008--1018}.
\newblock
\urldef\tempurl%
\url{https://www.jstor.org/stable/2779449}
\showURL{%
\tempurl}


\bibitem[White(1986)]%
        {White-1986}
\bibfield{author}{\bibinfo{person}{Michael~J. White}.} \bibinfo{year}{1986}\natexlab{}.
\newblock \showarticletitle{Segregation and Diversity Measures in Population Distribution}.
\newblock \bibinfo{journal}{\emph{Population Index}} \bibinfo{volume}{52}, \bibinfo{number}{2} (\bibinfo{year}{1986}), \bibinfo{pages}{198--221}.
\newblock
\urldef\tempurl%
\url{https://www.jstor.org/stable/3644339}
\showURL{%
\tempurl}


\end{thebibliography}
\end{document}